\global\def\draftcontrol{0}

\def\versionno{ time -- draft -- 11.13.03  }
\catcode`\@=11

\expandafter\ifx\csname draftcontrol\endcsname\relax\global\def\draftcontrol{0}
\fi

{\count255=\time\divide\count255 by 60
\xdef\hourmin{\number\count255}
\multiply\count255 by-60\advance\count255 by\time
\xdef\hourmin{\hourmin:\ifnum\count255<10 0\fi\the\count255}}
\def\draftdate{\number\month/\number\day/\number\year\ \ \ \hourmin }

\newcommand\makepapertitle{\par
  \begingroup
    \renewcommand\thefootnote{\@fnsymbol\c@footnote}%
    \def\@makefnmark{\rlap{\@textsuperscript{\normalfont\@thefnmark}}}%
    \long\def\@makefntext##1{\parindent 1em\noindent
            \hb@xt@1.8em{%
                \hss\@textsuperscript{\normalfont\@thefnmark}}##1}%
     \newpage
     \global\@topnum\z@   % Prevents figures from going at top of page.
     \@makepapertitle
     \thispagestyle{empty}\@thanks
  \endgroup
  \setcounter{footnote}{0}%
  \global\let\thanks\relax
  \global\let\makepapertitle\relax
  \global\let\@makepapertitle\relax
  \global\let\@thanks\@empty
  \global\let\@author\@empty
  \global\let\@date\@empty
  \global\let\@title\@empty
  \global\let\title\relax
  \global\let\author\relax
  \global\let\date\relax
  \global\let\and\relax
  \def\version{\let\version\@version\@gobble}
}
\def\@makepapertitle{%
  \newpage
   \ifnum\draftcontrol=1 {}
   \version\versionno
   \vskip 3em%
   \else
   \hfill\hbox to 3cm {\parbox{4cm}{\@pubnum}\hss}%
   \vskip 3em%
   \fi
   \begin{center}%
   \let \footnote \thanks
     {\LARGE \@title \par}%
     \vskip 1.5em%
     {\normalsize%\large
       \lineskip .5em%
       \begin{tabular}[t]{c}%
         \@author
       \end{tabular}\par}%
     \vskip 1em%
     {\@bstract}%
     \end{center}%
     \vskip .5em
     \@date%
   \par
}

\gdef\@pubnum{}
\def\pubnum#1{%
  \gdef\@pubnum{#1}}

\gdef\@bstract{}
\def\Abstract#1{%
  \gdef\@bstract{%
   \parbox{\textwidth-0pc}{%
   \centerline{\bf Abstract}\penalty1000
   \noindent%\abstractfont \baselineskip=12pt
   \renewcommand\baselinestretch{1.0}
   {#1}}}
}

\def\ps@paper{\let\@mkboth\@gobbletwo%
     \ifnum\draftcontrol=1
        \def\@oddfoot{\hbox to \textwidth{\tiny \versionno \hfil\tiny\draftdate}%
        \hskip -\textwidth \hbox to \textwidth{\hfil\rm\thepage\hfil}}%
     \else\def\@oddfoot{\hbox to \textwidth{\hfil\rm\thepage\hfil}}
     \fi
     \let\@evenfoot\@oddfoot
}
\def\body{\clearpage
          \pagestyle{paper}
        }
\newenvironment{acknowledgments}{%
\vskip 3.25ex
\noindent {\bf Acknowledgments}
}

\def\@version#1{\ifnum\draftcontrol=1
\typeout{}\typeout{#1}\typeout{}
\vskip3mm\centerline{\hbox{\fbox{\normalsize{\tt DRAFT -- #1 -- }
                   {\draftdate}}}}\vskip3mm
\fi}
\let\version\@version
\long\def\eqlabel#1{\ifnum\draftcontrol=1
                    \tag@false  % there are some problems with multline without this
                    \tag*{(\theequation) \hbox to -0.2cm{\hspace{0cm}\small{#1}\hss}}
                    \refstepcounter{equation} 
                    \edef\@currentlabel{\theequation}
                    \ltx@label{#1}          % use old LaTeX \label instead of new definition
                    \else
                    \label{#1}
                    \fi
                    }

\documentclass[12pt,letterpaper]{article}
\usepackage{amsmath,amssymb,array,calc,epsfig}
\ifnum\draftcontrol=1
\fi
\renewcommand\baselinestretch{1.25}
\renewcommand\section{\@startsection {section}{1}{\z@}%
                                   {-3.5ex \@plus -1ex \@minus -.2ex}%
                                   {2.3ex \@plus.2ex}%
                                   {\normalfont\large\bfseries}}
\renewcommand\subsection{\@startsection{subsection}{2}{\z@}%
                                     {-3.25ex\@plus -1ex \@minus -.2ex}%
                                     {1.5ex \@plus .2ex}%
                                     {\normalfont\normalsize\bfseries}}
\renewcommand\subsubsection{\@startsection{subsubsection}{3}{\z@}%
                                     {-3.25ex\@plus -1ex \@minus -.2ex}%
                                     {1.5ex \@plus .2ex}%
                                     {\normalfont\normalsize\it}}

\numberwithin{equation}{section}

\def\projective   {{\mathbb P}}

\def\be     {\begin{equation}}
\def\ende       {\end{equation}}

\def\revise#1       {\marginpar{\rule{2mm}{1cm} #1}}

\newcommand{\nc}{\newcommand}

\def\bea        {\begin{eqnarray}}
\def\eea        {\end{eqnarray}}
\nc{\e}{{\rm exp}}

\nc{\cosech}{{\rm cosech}}

\nc{\Li}{{\rm Li_{2}}}

\nc{\li}{\lambda_{i}}
\nc{\lj}{\lambda_{j}}
\nc{\lk}{\lambda_{k}}
\nc{\laml}{\lambda_{l}}

\nc{\mi}{\mu_{i}}
\nc{\mj}{\mu_{j}}
\nc{\mk}{\mu_{k}}
\nc{\ml}{\mu_{l}}

\nc{\om}{\omega}

\nc{\non}{\nonumber}

\catcode`\@=12

\begin{document}

\title{Chern-Simons matrix model for local conifold} 

\pubnum{%
hep-th/0409136}

\date{}

\author{Vadim Yasnov\footnote{yasnov@physics.usc.edu}\\[0.4cm]
\it Department of Physics and Astronomy\\
\it University of Southern California \\
\it Los Angeles, CA 90089, USA \\[0.2cm]
}

\Abstract{We check the large N transition proposed  by Aganagic and Vafa and further studied  by Diaconescu, Florea and Grassi using the large N limit of the corresponding Chern-Simons two matrix model. We find the spectral curve and calculate the genus zero free energy.
}

\enlargethispage{1.5cm}

\makepapertitle

\body

%------------------------------------------------------------------------------------------------------------------------
\section{Introduction}
The conifold transition is a duality between open and closed strings. In the topological $A$-model, this is a duality between the open string model on the deformed conifold  and the closed string model on the resolved conifold. This was originally studied by taking the partition function of large N Chern-Simons (CS) theory on $S^3$ expanded in a 't Hooft limit  and then summing over the holes to get a closed string theory \cite{Gopakumar:1998ii,Gopakumar:1998ki}. A worldsheet proof of this duality has since been provided \cite{Ooguri:2002gx} and generalized in \cite{Ooguri:04o}. 

It is interesting to extend this duality. One possible generalization is the local conifold, a Calabi-Yau (CY) that locally looks like a conifold but has an extra vertex in its toric diagram. The large N transition for this 
type of geometry was conjectured in \cite{av} and further investigated in \cite{dfg}. The all genus partition function for closed strings has been calculated using BPS invariants. The open string story is more complicated.
Since the A model geometry is not a cotangent bundle the Chern-Simons theory receives open string instanton corrections \cite{Witten:1992fb}. Nevertheless, the corresponding CS free energy has been calculated \cite{dfg} and
the Kahler parameters for open and closed strings have been related. An interesting half integral shift in the
second Kahler parameter is observed. We want to stress that since the shift is not seen at the tree level the only available calculation that confirms this shift is that of Diaconescu, Florea and Grassi.  A more simple alternative calculation would provide us with an additional insight. 

Fortunately, there is a way to write the CS 
partition function as a matrix model integral \cite{Marino:2002fk,Aganagic:2002wv}. In addition,
the mirror B model geometry can be obtained by taking the large N limit of the matrix model. The mirror to the local conifold is known, we will show that  
the spectral curve of the corresponding matrix model matches this B model geometry, thus providing  further 
evidence of the proposed duality. Moreover, the dictionary between open and closed string sides is
derived using not only the free energy but also geometry, importantly  
we  confirm the half integral shift found in \cite{dfg}. This is  the main result of the article.
We also calculate the genus 
zero free energy using the matrix model. The technique we use to get the spectral curve is inspired by \cite{km} where spectral 
curves for hermitian two matrix models are studied.

In the second section we give a concise description of the CY geometry from the A and B model sides. The third
section is devoted to the matrix model integral and the equations of motion. The spectral curve is found and discussed in section 4.      
\section{Geometry}
In this section we briefly introduce the underlying CY geometry. For details see \cite{av,dfg}.
We start with the deformed local conifold $Y$ in the A model. The toric diagram for $Y$ has an extra vertex compare to the toric diagram of the deformed conifold. N \mbox{D-b}ranes are wrapped on the lagrangian $S^3$ in this geometry.
The topological open string A model is  $U(N)$ CS on $S^3$ with open string instanton corrections \cite{dfg} that have an interpretation of non local Wilson loop effects in the CS theory \cite{Witten:1992fb}.   
The size of the $S^3$ is a complex structure moduli therefore A model amplitudes are insensitive to it. Taking
the $S^3$ to the zero size leads to a singular CY that locally looks like a conifold. The singularity can be 
resolved by blowing up a $\projective^1$.  
\begin{figure}[bth!]
\centerline{ \epsfig{file=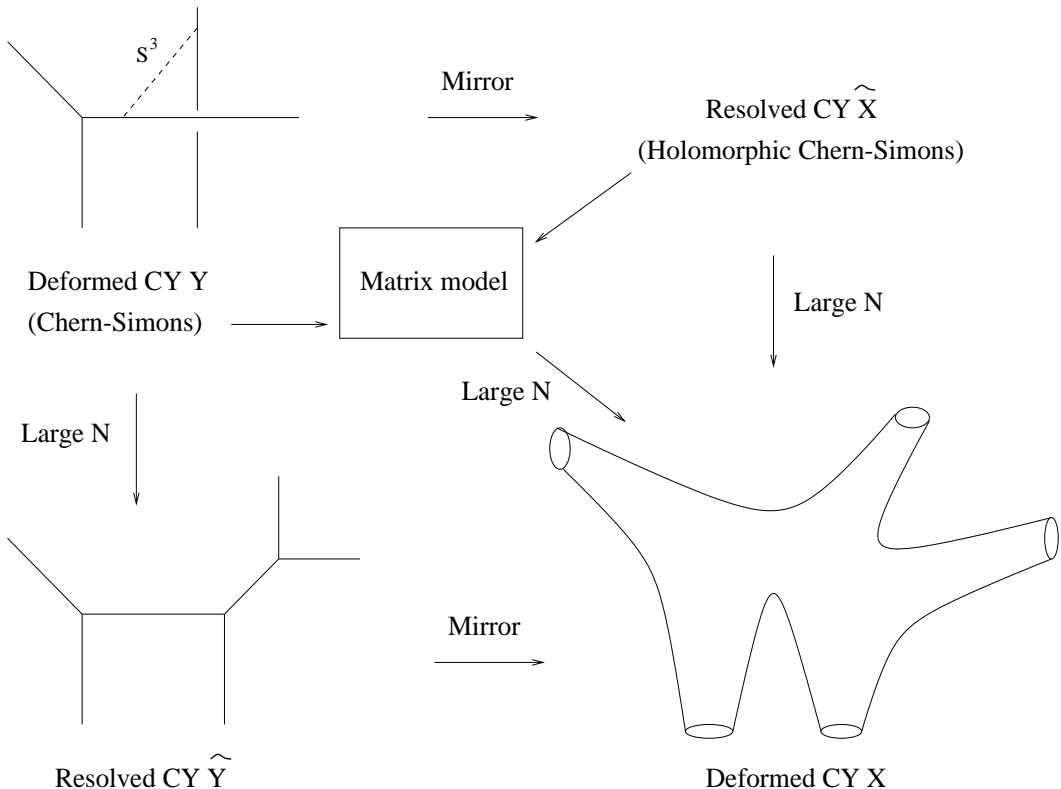,width=14cm,height=10.4cm}}
\caption{The web of large N dualities and mirror symmetries.} 
\label{web}
\end{figure}
The geometric transition maps the open topological model on the deformed local conifold $Y$ to the closed  string model on the resolved CY $\tilde Y$ that has two Kahler parameters, $\tilde t$ and $\tilde s$, the sizes of the 
two $\projective^1$'s. 
In terms of the gauge theory the transition leads to the large N limit of the CS theory. 
If embedded in the full string theory this geometric transition describes  dissolving N D6-branes wrapped on the vanishing $S^3$ and replacing them by a flux. This large N duality was conjecture by Aganagic and Vafa in \cite{av}
and further studied by Diaconescu, Florea and Grassi in \cite{dfg}. 

There is the mirror side of the story. The mirror to $Y$ is the resolution of the following 
singular CY \cite{Hori:2000kt,Hori:00iv}
\be
\label{singCY}
xz=(e^{u-t}-1)(e^u+e^v-1).
\ende
The singularity gets resolved by blowing up a holomorphic $\projective^1$ at the points where the r.h.s. of 
equation (\ref{singCY}) is zero.  Let denote this CY as $\tilde X$.
The branes mirror to the A model branes wrap this $\projective^1$. The effective topological
open string theory is holomorphic Chern-Simons \cite{Witten:1992fb}. The large N transition for this geometry
closes the web of mirrors and large N transitions giving us the mirror to $\tilde Y$, the A model resolved CY. The
mirror of $\tilde Y$ is the deformed CY $X$
\be
\label{scurve}
xz=(1-e^{-\hat u})(1-e^{\hat u-\hat t})-e^{-\hat v}(1-e^{\hat u-\hat t-\hat s})
\ende
which is written in the flat coordinates \cite{Aganagic:01kv}.
The web of the mirror maps and the large N transitions is summarised on {fig.\ref{web}} 

The genus zero $\hat s$ dependent part of the free energy for closed strings in the resolved geometry $\tilde Y$ was found in \cite{av} by computing classical periods in the mirror B model geometry $X$. It was compared with the disk computations for open strings. Furthermore, the $\hat s$ dependent part of the all genus partition function for closed strings was derived using BPS invariants. The $\hat t$ dependent part of the all genus free energy was obtain in \cite{dfg} by compactifying the model and calculating BPS invariants. The open string computations are more complicated since they involve the calculation of non local Wilson loops in the gauge theory. 
The open string partition function was obtained in \cite{dfg} and matched to the closed string partition function.  
The dictionary between the open and closed string parameters found by Diaconescu, Florea and Grassi,  
\be
\label{shiftdfg}
\hat s=-g_sN\hspace{2cm}\hat t=t_{op}+\frac{g_sN}{2},
\ende
contains the curious half integral shift in the second Kahler parameter $\hat t$. This shift is not seen   
at the tree level. The shift is interpreted as a contribution of degenerate open string instantons. So far
this is the only available open string computation that confirms this shift.

An alternative computation of the CS partition function can be done by using matrix models. Moreover,
the spectral curve of the matrix model is a part of the mirror to the large N dual geometry.
In this context, CS matrix models are a powerful tool for studying relationship between
closed and open string parameters.
The corresponding matrix models were introduced in \cite{Aganagic:2002wv}. They differ from more familiar hermitian
models by the measure factor. The integration goes over the unitary group. CS matrix models have been used to study the large N dualities for the conifold \cite{Aganagic:2002wv}, the orbifold of the conifold \cite{Halmagyi:2003fy} and the lens spaces \cite{Halmagyi:2003ze,lens}.

In this paper we find the spectral curve of the CS matrix model for the local conifold and confirm eq. (\ref{shiftdfg}).

\section{Matrix model}
The local conifold under consideration contains the lagrangian $S^3$ over which N \mbox{D-b}ranes are wrapped. Since the 
geometry is not a cotangent bundle the CS theory gets modified by open string instanton corrections    \cite{Witten:1992fb}. The corresponding two matrix model for the CS theory 
has been proposed in \cite{Aganagic:2002wv} and reads
\be
\label{Z}
Z\sim\int\prod_i^N du_i dv_i\Delta(u)\Delta(v)exp\left (-\frac{1}{g_s}\sum_i^N (u_iv_i+V_1(v_i)+V_2(u_i))\right ),
\ende
where the measure factor is
\be
\label{measure}
\Delta (x)=\prod_{i<j}2\sinh \left (\frac{x_i-x_j}{2}\right )
\ende
and the potentials are given by
\be
\label{v1}
V_1(v)=tv,
\ende
\be
\label{v2}
V_2(u)=-\sum_{n=1}^\infty\frac{1}{n^2}e^{nu}.
\ende
The integration measure is the Haar measure on the space of two commuting unitary matrixes. The two extrema of the matrix model action correspond to the north and south poles of the holomorphic $\projective^1$ in the blown up B model geometry (\ref{singCY}). That is a characteristic feature of CS matrix models, their potentials encode the mirror B model geometry. It has a simple explanation since CS matrix models can also be seen as 
coming from holomorphic CS on the B model side. 

The matrix model has two sets of eigenvalues. They interact only through the linear term $\sum_iu_iv_i$.
This gives a hint that one should look at a similar hermitian two matrix model which has two matrixes
$X$ and $Y$ that interact though ${\rm Tr}XY$ in the matrix model action. There is a vast literature on the 
subject, see for example \cite{2mm}. The one of a particular interest for us is the paper by Kazakov and Marshakov \cite{km} where a simple method for finding spectral curves for hermitian two matrix models is proposed. The model also has applications to 2D string theory \cite{km_cont}.
   
Anticipating taking the large N limit denote $s=g_sN$ as the 't Hooft parameter. Let introduce two resolvents, one for each set of eigenvalues, 
\be
\label{om1}
\omega_1(v)=g_s\sum_i\coth\left (\frac{v-v_i}{2}\right ),
\ende
\be
\label{om2}
\omega_2(u)=g_s\sum_i\coth\left (\frac{u-u_i}{2}\right ).
\ende
They enjoy the following boundary conditions 
\be
\label{bc}
\lim_{x\rightarrow\pm\infty}\omega_{1,2}(x)=\pm s.
\ende
The equations of motion (EM) can be conveniently written in terms of the resolvents as
\be
\label{veq}
u_i=t+\frac{1}{2}\omega_1(v_i),
\ende
\be
\label{ueq}
v_i=\ln (1-e^{u_i})+\frac{1}{2}\omega_2(u_i).
\ende 
There is another form of EM that resembles a hermitian two matrix model. 
Let $U=e^u$, $V=e^v$, ${\rm W}_1(V)=e^{\omega_1(V)/2}$ and ${\rm W}_2(U)=e^{\omega_2(U)/2}$ then
\be
\label{UofV}
U=e^t{\rm W}_1(V),
\ende
\be
\label{VofU}
V=(1-U){\rm W}_2(U).
\ende
Our goal is to take the large N limit of the above equations and derive the spectral curve.

\section{Spectral Curve}
In this section we consider the large N limit of the CS matrix model and find the corresponding 
spectral curve. The spectral curve of the matrix model is a part of the large N dual B model geometry. The goal is to compare the spectral curve of the matrix model to $X$, the mirror to the conjectured large N dual $\tilde Y$. 
  
We want to proceed along the lines of \cite{km} by Kazakov and Marshakov, where the spectral curve for the hermitian two matrix model is derived. Although hermitian models are quite different from CS matrix models there is an obvious analogy between their EM observed in (\ref{UofV}) and (\ref{VofU}).
  
In the large N limit the  eigenvalue distribution  becomes a continuous function and poles of the 
resolvents spread into the cuts. Since $U$ and $V$ in the EM in the large N limit lie on some cuts
the functions ${\rm W}_1(U)$ and ${\rm W}_2(V)$ in (\ref{UofV}),(\ref{VofU}) are understood as principal values. The EM  determine the behavior of the functions $U(V)$ and $V(U)$ at infinity and at zero . In addition, as explained  in  \cite{km} these two functions in the large N limit are inverse of each other 
\be
\label{inverse}
U(V(\tilde U))=\tilde U.
\ende
Furthermore, they completely fix the spectral  curve of the model. Either one of them is actually the spectral curve. The functions $U(V)$ and $V(U)$ have the following behavior at infinity 
\be
\label{UofVin}
U(V)\sim e^{t+s/2}+O(1/V),
\ende
\be
\label{VofUin}
V(U)\sim -Ue^{s/2}+O(U^0)
\ende
and at zero
\be
\label{UofV0}
U(V)\sim e^{t-s/2}+O(V),
\ende
\be
\label{VofU0}
V(U)\sim e^{-s/2}+O(U).
\ende
To find the spectral curve we are going to apply the method proposed by Kazakov and Marshakov for hermitian two matrix models. It turns out that our case when $U(V)$ goes to a constant at infinity is more subtle then the general one. So we 
start with the general case and briefly review the technique.  We suppose that $U(V)\sim V^n$ for big $V$ and $V(U)\sim U^m$ for big $U$. Then the spectral curve is given by 
\be
\label{FofUVnm}
F(U,V)=U^{m+1}+V^{n+1}-V^nU^m+...=0,
\ende
where the function $F(U,V)$ is polynomial and only the first three terms with the highest powers of $U$ and $V$ are written explicitly. The functional form of $F(U,V)$ is completely fixed by the given behavior at infinity. The coefficients of the lower power terms can be found using the known behavior of $U(V)$ and $V(U)$ at zero and by fixing certain period integrals of the spectral curve.   Note that if, for example, $n=0$, terms of the form 
$U^k-U^{k-lm}+V^l$ in $F(U,V)$ for any integers $k$ and $l$ are completely consistent with the given asymptotic at 
infinity. It is clear that for this degenerate case one can not determine the functional form of the spectral curve just by looking at the functions $U(V)$ and $V(U)$ at infinity. However, it is reasonable to consider the matrix model (\ref{Z}) as a smooth limit of  a matrix model
with a potential that has higher powers of $V$. Let $\tilde V_1(v)$ be such that 
\be
{\tilde V'}_1(v)=\ln \left (e^t-e^rV^2\right ).
\ende
When $r\rightarrow -\infty$ the above potential reduces to $V_1(v)$. With this new potential \mbox{$U(V)\sim V^2$}
 at infinity and so the formula (\ref{FofUVnm}) with $n=2$ and $m=1$ can be applied leading to 
\be
F(U,V)=\alpha V^3+\beta U^2+\gamma V^2 U +\delta V^2 +\epsilon U V+\varepsilon U+\zeta V+1.
\ende
The next step is to see what coefficients survive taking $r\rightarrow -\infty$ limit. As it is shown in the appendix the spectral curve for the cubic potential looses three terms in this limit and takes the form
\be
\label{FofUVab}
F(U,V)=\beta U^2+\epsilon U V+ \varepsilon U +\zeta V+1=0.
\ende 
There are four coefficients that can be found by the known behavior at infinity and at zero. Carrying on this 
calculation one gets 
\be
\label{FofUV}
F(U,V)=U^2+e^{-s/2} UV-e^tU-(1+e^{t-s/2})V+e^{t-s/2}=0.
\ende
The simplest check of the above expression is to take the limit $s=0$. In this limit  all eigenvalues 
must be located at the classical extremum of the matrix model action. That is indeed true. The spectral curve in the classical limit splits  into the two independent equations $V=e^t$ and $V=1-U$ that are precisely the location 
of the holomorphic $\projective^1$ over which the branes are wrapped.
The spectral curve also gives us the resolvents
\be
\label{W1}
\hspace{-0.3cm}
{\rm W}_1(V)=\frac{e^{-t}}{2}\left (1+e^{t-s/2}-e^{-s/2} V+\sqrt{{(e^{-s} V-1-e^{t-s/2})}^2+4e^tV-4e^{t-s/2}}\right ),
\ende
\be
\label{W2}
{\rm W}_2(U)=e^{s/2}\frac{U-e^{t-s/2}}{U-e^{t+s/2}},
\ende
where the plus sign in front of the square root  is chosen in order to have the correct limit at infinity. Note that
the point $v=-\infty$ where the resolvent $\omega_1(v)$ is equal to $-s$ is located on the other sheet of the spectral curve
which corresponds to the minus sign before the square root in (\ref{W1}). In the next section we calculate the genus zero
free energy as an integral over $u$ of the spectral curve so we also need the expression for the spectral curve in 
the $u$ variable
\be
\label{vofu}
v(u)=\ln\frac{(1-e^u)(e^u-e^{t-s/2})}{e^{-s}e^u-e^t}.
\ende
\begin{figure}[bth!]
\centerline{\epsfig{file=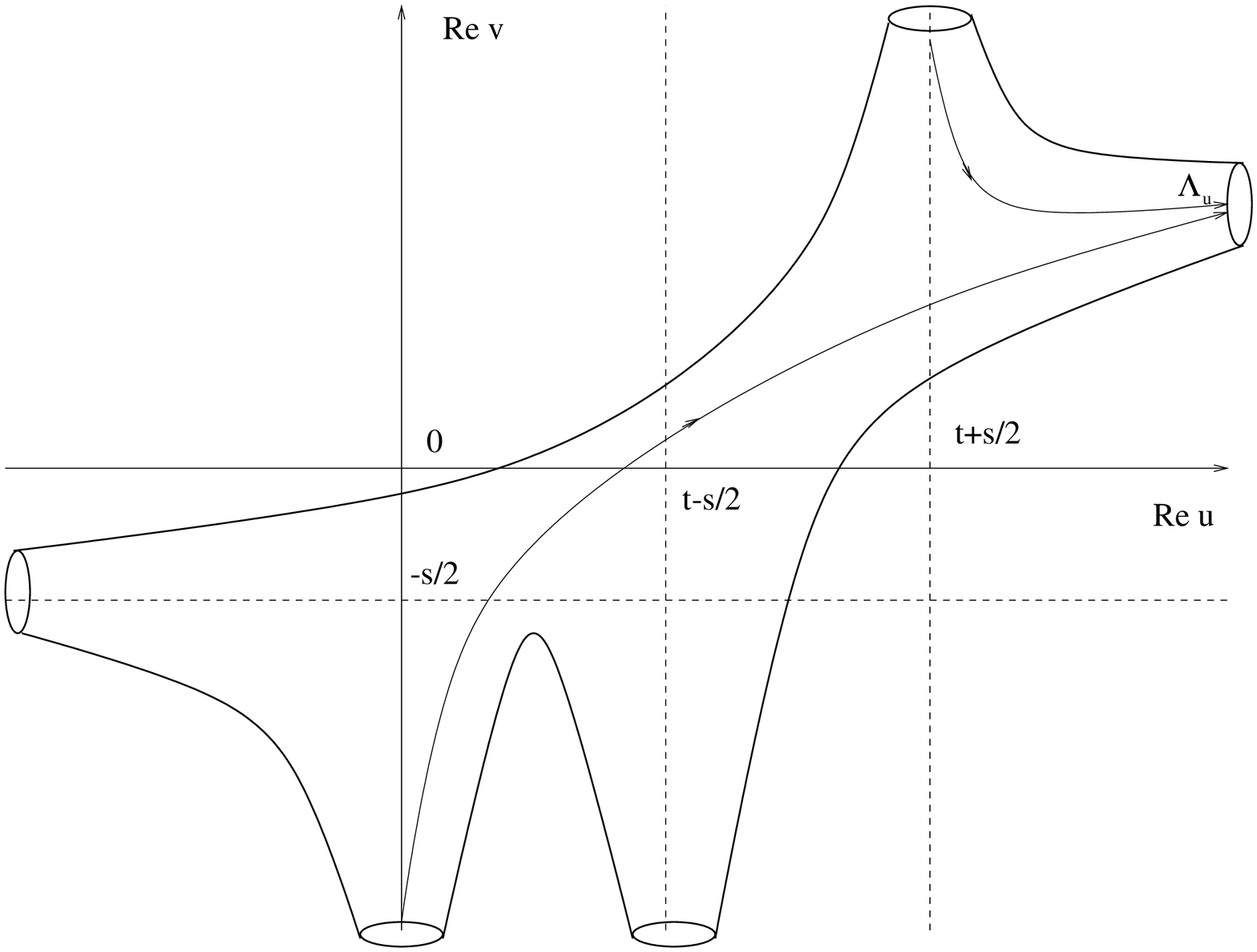,width=14cm,height=10.4cm}}
\caption{The spectral curve for the CS matrix model. The two contours are the noncompact cycles used in the calculation of 
the genus zero free energy.} 
\label{kri}
\end{figure}
We now compare our spectral curve to the spectral curve coming from CS on $S^3$ or the A model conifold    \cite{Aganagic:2002wv}. Since the toric
diagram for the resolved A model geometry has an extra vertex  we expect an extra cylindrical sheet in the mirror B model. Let look at the function $u(v)=t+\ln{\rm W}_1(v)$. The corresponding Riemann surface 
consists of the three cylindrical sheets. The physical sheet is the one where the resolvent $\omega_1(v)$ is 
finite at  infinity. The physical sheet is glued to the second cylinder along the square root cut. On this second
sheet the resolvent $\omega_1(v)$ has the minus sign in front of the square root and is equal to $-s$ at $v=-\infty$. The second sheet has the logarithmic cut that starts at $v=-s/2$ and goes to infinity. The logarithmic cut connects
the second sheet to the third one that does not have any square root cuts. The  logarithmic cut is responsible for 
the extra vertex in the A model toric diagram that differs it from the conifold. The square root cut shrinks to the zero size when $s=0$ while the logarithmic cut is present even in the classical limit although the location of the branch point gets modified by quantum corrections. The spectral curve is depicted on fig.\ref{kri}.

It is easy to see that the spectral curve looks very similar to (\ref{scurve}) that is the mirror to the conjectured large N dual of $Y$. It turns out that  there is a simple coordinate transformation
\be
\label{vvhat}
v=-\hat v +s/2
\ende
\be
\label{uuhat}
u=-\hat u
\ende
that brings the spectral curve to the form
\be
\label{scurvets}
e^{t-s/2} e^{2\hat u +v}-e^{t+s/2}e^{2\hat u}-(1+e^{t-s/2})e^{\hat u \hat v}+e^{\hat v}+e^{\hat u}=0
\ende
which is exactly the non trivial part of the deformed B model CY $X$ in (\ref{scurve}). Now it is easy to read off
the relationship between the open and closed string Kahler parameters
\be
\label{dictionary}
\hat s =-s\hspace{2cm}\hat t =-t+s/2,
\ende
where $\hat t$ and $\hat s$ are the closed string flat coordinates. Since the open string Kahler parameter $t$ that is used in the matrix model is equal to $-t_{op}$ of \cite{dfg} the above equations are precisely those
found by Diaconescu, Florea and Grassi (\ref{shiftdfg}). It confirms the large N conjecture and the  shift by $s/2$ in the second Kahler parameter.

Therefore we have derived the mirror to the large N dual of the local conifold. It coincides with one proposed by
Aganagic and Vafa in \cite{av}. This is a strong evidence that the large N conjecture is indeed true. 

To further study the large N transition  one has to find the matrix model free energy and compare it with
the closed string calculation. 

\section{Free energy}
In this section we calculate the genus zero free energy. As for hermitian matrix models the derivatives of the 
free energy with respect to $t$ and $s$ can be written as integrals of the resolvents over some noncompact cycles. 
A useful relationship that we are going to explore through this section is \cite{Halmagyi:2003ze}
\be
\label{logtoomega}
g_s\sum_i\ln\left(2\sinh \frac{x-x_i}{2}\right )=-\frac{1}{2}P\int^\Lambda_x\omega (z)dz-\frac{1}{2}\sum_i x_i,
\ende
where $\Lambda$ is a point at infinity on the physical sheet of the resolvent and $x$ is a point on the cut. All $\Lambda$ dependent terms are omitted. 
The sum $g_s\sum_iv_i$ is very important and represents a period integral, meanwhile the sum $g_s\sum_iu_i=ts$ is
trivial and will be omitted.    
Since we have to sum over both $u$ and $v$ eigenvalues it is convenient to
introduce $\Lambda_u$ and $\Lambda_v$, points on the physical sheet at $u$ and $v$ infinities respectively. 

It immediately follows that the derivative of the genus zero free energy ${\rm F}_0$ with respect to $t$ can be cast into the form of the integral of the spectral curve
\be
\label{dtf0int}
\partial_t{\rm F}_0 =g_s\sum_i v_i=g_s\sum_i\ln (1-e^{u_i})=-\int^{\Lambda_u}_0v(u)du.
\ende
Here, the summed over $i$ EM have been used and all polynomial in $t$ and $s$ terms are skipped. $v(u)$ is given by
(\ref{vofu}). The integral can be easily evaluated leading to 
\be
\label{dtf0}
\partial_t{\rm F}_0=\sum^\infty_{n=1}\frac{1}{n^2}\left (e^{n(t-s/2)}-e^{n(t+s/2)} \right ).
\ende
The derivative of the free energy with respect to the 't Hooft parameter is the variation of the matrix model
action due to eigenvalues $u$ and $v(u)$ added to the square root cut
\be
\label{dsF0a}
\partial_s{\rm F}_0=-uv+V_1(v)+V_2(u)+g_s\sum_i\ln\left (2\sinh\frac{v-v_i}{2}\right )+g_s\sum_i\ln\left (2\sinh\frac{v-v_i}{2}\right ).
\ende
The two last sums in the above equation can be converted to the integrals of the resolvents by means of (\ref{logtoomega}). Using (\ref{dtf0int}) the expression can be further transformed to the form
\be
\label{dsF0b}
\partial_s{\rm F}_0=-uv-\int^{\Lambda_v}_vu(v)dv-\int^{\Lambda_u}_u v(u)du+\frac{1}{2}\int^{\Lambda_u}_0v(u)du.
\ende
The last step is to use that $u$ and $v$ are related by the spectral curve (\ref{FofUV}). This takes us to (see fig.\ref{kri}) 
\be
\label{dsF0int}
\partial_s{\rm F}_0=-\int^{\Lambda_u}_{s/2+t}v(u)du+\frac{1}{2}\int^{\Lambda_u}_0v(u)du
\ende
with the result of the integration being
\be
\label{dsF0final}
\partial_s{\rm F}_0=-\sum_{n=1}^\infty\frac{1}{n^2}\left (e^{ns}+\frac{1}{2}e^{n(t-s/2)}+\frac{1}{2}e^{n(t+s/2)}\right ).
\ende
The energy becomes
\be
{\rm F}_0=-\sum^\infty_{n=1}\frac{1}{n^3} \left (e^{ns}+e^{n(t+s/2)}-e^{n(t-s/2)}\right ).
\ende
With the identifications (\ref{dictionary}) this free energy matches the closed string free energy found in  \cite{dfg}. 

The fact that $\partial_s{\rm F}_0$ is the combination of the integrals over the two independent   cycles is a consequence of
the half integral shift in (\ref{dictionary}). The open string flat coordinate $t$ is not a flat coordinate  for 
the closed string model. In the matrix model language if we interpret the matrix eigenvalues as D-branes the shift occurs because the anti D-branes at $u=0$ form a bound state with the D-branes spread between $u=t-s/2$ and $u=t+s/2$. Moving two D-branes to infinity must be supplemented  with  pulling one anti D-brane to infinity.

It leads to another check of the large N transition. The free energy calculated from the open string prospectives 
is the same as the closed string energy.   
  
\section*{Conclusion}
We have provided a strong evidence that the conjectured large N transition is indeed true. We see that the second
Kahler parameter $\hat t$ does receive the half integral shift that supposedly comes from the degenerate open
string instantons. 

We have found the method of calculating spectral curves for the wide class of two matrix Chern-Simons models.
It is easy to see that the method works for any polynomial (after exponentiating ) potentials. The spectral curve 
is given by (\ref{FofUVnm}). 

It is interesting to consider more general Chern-Simons matrix models, for example, the matrix models for the geometries considered in \cite{dfg2}. 
Given a relative simplicity of  matrix model calculations it is conceivable that
Chern-Simons matrix models can provide many new interesting examples of large N transitions. 

Another intriguing question is to relate universal correlators for 3D partitions in CY crystals to CS matrix 
models \cite{Dijkgraaf:04st}.  

\begin{acknowledgments}
I would like to thank Nick Halmagyi for the discussion and reading the manuscript. The work was supported by
a DOE grant DE-FG03-84ER40168.  
\end{acknowledgments} 

\begin{appendix}
\section{From cubic to linear potential}
The new EM with the cubic potential are
\be
U(V)=(e^t-e^rV^2){\rm W}_1(V),
\ende
\be
V(U)=(1-U){\rm W}_2(U),
\ende
where parameter $r$ is intended to be taken to minus infinity at the end. Our concern is only the behavior of $U(V)$ and 
$V(U)$ at infinity which is given by
\be
\label{UofVinr}
U(V)\sim -e^re^{s/2}V^2-e^rAV+(e^{t+s/2}-e^rB)+O(1/V),
\ende
\be
\label{VofUinr}
V(U)\sim -Ue^{s/2}+O(U^0).
\ende
Here $A$ and $B$ stand for the expansion coefficients of the resolvent ${\rm W}_1(V)$. The resolvent has to be expanded 
up to $1/V^2$ in order to keep the constant term in (\ref{UofVinr}). The later is needed since after putting
$e^r$ to zero eq. (\ref{UofVinr}) must reduce to (\ref{UofVin}). The spectral curve that is consistent with the above behavior at infinity is of the form \cite{km}
\be
\alpha V^3+\beta U^2+\gamma V^2 U +\delta V^2 +\epsilon U V+\varepsilon U+\zeta V+1=0.
\ende 
We want to find what coefficients in the above equations are zero if $e^r$ is zero. As $V$ goes to infinity we have
to zero all terms of order $V^4$, $V^3$ and $V^2$ which gives
\be
e^{r+s/2}\beta-\gamma =0,
\ende
\be
\alpha +2 \beta e^{2r+s/2} A-\gamma e^rA-\epsilon e^{r+s/2}=0,
\ende  
\be
e^{2r}A^2\beta -2\beta (e^{t+s/2}-e^r B)e^{r+s/2}+\gamma (e^{t+s/2}-e^r B)+\delta-\epsilon e^r A=0.
\ende
As $U$ is taken to infinity only terms of order $U^3$ have to be kept
\be
-\alpha e^{s/2}+\gamma =0.
\ende
Clearly, when $e^r=0$ one must have $\alpha =\gamma =\delta =0$. Therefore, the appropriate functional form of 
the spectral curve with the linear potential is 
\be
\beta U^2+\epsilon U V+ \varepsilon U +\zeta V+1=0.
\ende
\end{appendix}

\end{document}